\documentclass[journal]{IEEEtran}
\usepackage{mathrsfs}
\usepackage{pifont}
\usepackage{bbding}
\usepackage{tipa}
\usepackage{amsmath,epsfig}
\usepackage{amssymb}
\usepackage{amsfonts}
\usepackage{graphicx}
\ifCLASSINFOpdf
\else
\fi
\begin{document}
\newtheorem{theorem}{Theorem}
\newtheorem{proposition}{Proposition}
\newtheorem{definition}{Definition}
\newtheorem{lemma}{Lemma}
\newtheorem{corollary}{Corollary}
\newtheorem{remark}{Remark}
\newtheorem{construction}{Construction}

\newcommand{\supp}{\mathop{\rm supp}}
\newcommand{\sinc}{\mathop{\rm sinc}}
\newcommand{\spann}{\mathop{\rm span}}
\newcommand{\essinf}{\mathop{\rm ess\,inf}}
\newcommand{\esssup}{\mathop{\rm ess\,sup}}
\newcommand{\Lip}{\rm Lip}
\newcommand{\sign}{\mathop{\rm sign}}
\newcommand{\osc}{\mathop{\rm osc}}
\newcommand{\R}{{\mathbb{R}}}
\newcommand{\Z}{{\mathbb{Z}}}
\newcommand{\C}{{\mathbb{C}}}
%

\title{PAPR Reduction Method Based on Parametric Minimum Cross Entropy for OFDM Signals}
\author{{Yajun~Wang, Wen~Chen,~\IEEEmembership{Member,~IEEE}, and
Chintha Tellambura,~\IEEEmembership{Senier Member,~IEEE}}
\thanks{Manuscript received November 1, 2009. The associate editor coordinating
the review of this letter and approving it for publication was Jaap
van de Beek.}
\thanks{Yajun~Wang and Wen~Chen are with Department of Electronic Engineering,
Shanghai Jiaotong University, Shanghai, 200240 PRC. Yajun Wang is
also with the State Key Laboratory of Integrated Services Networks,
and Wen~Chen is also with SEU SKL for mobile communications. e-mail:
\{wangyj1859, wenchen\}@sjtu.edu.cn.}
\thanks{Chintha Tellambura is with Department of Electrical and Computer
Engineering, University of Alberta, Edmonton, Canada,  T6G 2V4.
e-mail: chintha@ece.ualberta.ca.}
\thanks{This work is supported by NSF China \#60972031, by SEU SKL project \#W2000907,
and by national 973 project \#YJCB2009024WL.}}


%

\markboth{IEEE Communications Letters, 2010}{Shell
\MakeLowercase{\textit{et al.}}: Bare Demo of IEEEtran.cls for
Journals}
\maketitle

\begin{abstract}
The partial transmit sequence (PTS) technique has received much
attention in reducing the high peak to average power ratio (PAPR) of
orthogonal frequency division multiplexing (OFDM) signals. However,
the PTS technique requires an exhaustive search of all combinations
of the allowed phase factors, and the search complexity increases
exponentially with the number of sub-blocks.
 In this paper, a novel method based on parametric minimum cross entropy
(PMCE) is proposed to search the optimal combination of phase
factors. The PMCE algorithm not only reduces  the PAPR significantly,
but also decreases  the
computational complexity.  The
simulation results show that it  achieves more or less the same PAPR reduction as that of exhaustive search.
\end{abstract}



%
\IEEEpeerreviewmaketitle

\section{Introduction}\label{sec:1}
In various high-speed wireless communication systems,  orthogonal
frequency division multiplexing (OFDM) has been used widely due to
its inherent robustness against multipath fading  and resistance to
narrowband interference~\cite{IEEEconf:1}. However, one of the major
drawbacks of OFDM signals is the high peak to average power ratio
(PAPR) of the transmitted signal. Several solutions have been
proposed in recent years, such as clipping~\cite{IEEEconf:2},
coding~\cite{IEEEconf:3}, selected mapping (SLM)~\cite{IEEEconf:4},
partial transmit sequence (PTS)~\cite{IEEEconf:5} and
others~\cite{IEEEconf:6}. The  PTS~\cite{IEEEconf:5} technique is a
distortionless technique based on combining signal subblocks which
are phase-shifted by constant phase factors, which can reduce PAPR
sufficiently. But the exhaustive search complexity of the optimal
phase combination in PTS increases exponentially with the number of
sub-blocks. Thus many suboptimal PTS techniques have been developed.
the iterative flipping   PTS (IPTS) in~\cite{IEEEconf:7} has
computational complexity linearly proportional to the number of
subblocks. A neighborhood search is proposed in~\cite{IEEEconf:8} by
using gradient descent search. A suboptimal method
in~\cite{IEEEconf:9} is developed by modifying the problem into an
equivalent problem of minimizing the sum of the phase-rotated
vectors.

In this paper, we propose a novel phase optimization scheme, which
can efficiently reduce the PAPR of the  OFDM signals,  based on  the
parametric minimum cross entropy (PMCE) method~\cite{IEEEconf:11}.
The proposed scheme can search for  the nearly optimal combination
of the initial phase factors. The simulation results show that this
scheme can achieve a  superior PAPR reduction performance, while
requiring  far less  computational complexity than the existing
techniques including the cross entropy approach~\cite{IEEEconf:13}.

\section{OFDM System And PAPR}\label{sec:2}
In an OFDM system, a high-rate data stream is split into $N$
low-rate streams
 transmitted simultaneously by subcarriers. Each of the subcarriers is
 independently modulated by  using  a typical modulation scheme such
 as  phase-shift keying (PSK) or quadrature amplitude modulation
  (QAM). The  inverse discrete Fourier transform (IDFT)  generates
  the ready-to-transmit  OFDM signal.
For an input OFDM block $\textbf{X}=[X_0,\dots,X_{N-1}]^T$, where
$N$ is  the number of subcarriers, the discrete-time baseband
OFDM  signal $x(k)$ can therefore  be expressed as
\begin{equation}\label{eq1}
x(k)=\frac{1}{\sqrt{N}}\sum_{n=0}^{N-1}X_n e^{\frac{j2\pi
nk}{LN}},k=0,1,\cdots,LN-1,
\end{equation}
where $L$ is the oversampling factor. It was  shown
in~\cite{IEEEconf:10} that the oversampling factor $L=4$ is enough
to provide a sufficiently accurate estimate of the PAPR of OFDM
signals.

 The PAPR of $x(k)$ is defined as the ratio of the maximum
instantaneous power to the average power;  that is
\begin{equation}\label{eq2}
PAPR=\frac{\underset{0\leq n<LN}{\max}|x(k)|^2}{E[|x(k)|^2]}.
\end{equation}

\section{PTS Techniques}\label{sec:3}
\begin{figure}
\centering
\includegraphics[width=3.5in,angle=0]{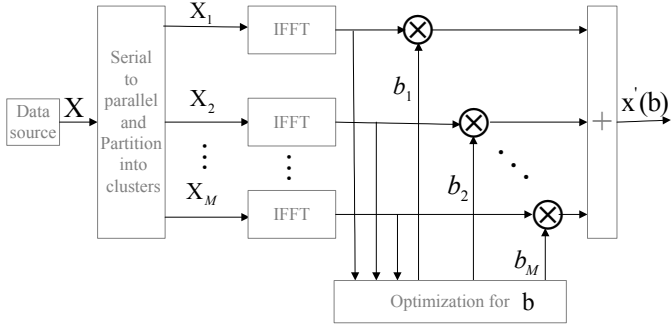}
\caption{Block diagram of the PTS technique.} \label{fig1}
\end{figure}

The  structure of the  PTS method is shown in Fig.~\ref{fig1}.
     The input data block $\textbf{X}$ is partitioned into $M$ disjoint
    sub-blocks $\textbf{X}_m,m=1,2,\dots M$ such that
    $\textbf{X}=\sum\limits_{m=1}^M\textbf{X}_m$. The
    sub-blocks are combined in the time domain  to minimize the
    PAPR. The  $L$-times oversampled time-domain   signal of $\textbf{X}_m$ is
denoted as $\textbf{x}_m,m=1,2,\dots M$, which are obtained by
taking an IDFT of length $NL$ on $\textbf{X}_m$ concatenated with
$(L-1)N$ zeros. Each $\textbf{x}_m$ is multiplied by a phase-weighting
factor $b_m=e^{j\phi_m}$, where~$\phi_m\in [0,2\pi)$~for
~$m=1,2,\dots M $. The goal of the PTS approach is to find an
optimal phase-weighted combination to minimize the PAPR. The
combined  transmitted signal in the time domain  can then  be
expressed as
\begin{equation}\label{eq3}
\textbf{x}^{'}(\textbf{b})=\sum_{i=1}^{M}b_i\textbf{x}_i,
\end{equation}
where
$\textbf{x}^{'}(\textbf{b})=[x_{1}^{'}(\textbf{b}),x_{2}^{'}(\textbf{b}),\cdots,x_{NL}^{'}(\textbf{b})]$.

In general, the selection of the phase factors is limited to a set
with a  finite number of elements to reduce the search complexity. The
set of allowed phase factors is
\begin{equation}\label{eq4}
\textbf{P}=\{e^{j2\pi\ell/W}|\ell=0,1,\dots,W-1\},
\end{equation}
where $W$ is the number of allowed phase factors. Thus,  $W^{M}$ sets of
phase factors are searched for  the optimal set of phase factors.
The search complexity increases exponentially with $M$, the number
of sub-blocks.

\section{Minimize PAPR Using Parametric Minimum
Cross Entropy (PMCE) Method }

The Parametric Minimum Cross Entropy Method
(PMCE) was first proposed by Rubinstein~\cite{IEEEconf:11} to solve rare event probability
estimation and counting problems. It is a parametric method to solve the well
known Kullback Minimum Cross Entropy (MinxEnt) problem~\cite{IEEEconf:12}.
The PMCE algorithm first casts a
deterministic optimization problem into an associate rare-event probability estimation, then solves  the resulting  program to obtain an optimally marginal distributions derived from
the optimal joint MinxEnt distribution.
This method finds the optimal parameters of the importance sampling distribution to efficiently estimate
 the desired quantity. For an accurate understanding
of  PMCE, the readers are referred to~\cite{IEEEconf:11}.


The minimum PAPR for PTS method is relative to the following  problem:
\begin{equation}\label{eq5}
\begin{split}
\text{Minimize }F(\textbf{b})=
&\frac{\max|x^{'}(\textbf{b})|^2}{E[|x^{'}(\textbf{b})|^2]},
\\s.t.\qquad &\textbf{b}\in\{e^{j\phi_m}\}^{M},
\end{split}
\end{equation}
where $\phi_m\in \{\frac{2\pi k}{W}|k=0,1,\dots,{W-1}\}$.
 The phase factor $\textbf{b}=\{-1,1\}^M$ is chosen in this paper and
 generated by using
 $\textbf{b}=1-2\textbf{c}$ from a binary vector
 $\textbf{c}=\{c_i\}_{i=0}^{M-1}$.
 Thus minimization of (\ref{eq5}) is translated into the following problem:
\begin{equation}\label{eq6}
\begin{split}
\text{Minimize } F(\textbf{c})=
&\frac{\max|x^{'}(1-2\textbf{c})|^2}{E[|x^{'}(1-2\textbf{c})|^2]},
\\s.t.\qquad &\textbf{c} \in \{0,1\}^{M}.
\end{split}
\end{equation}
%
%
%
Each element of $\textbf{c}$ can be modeled as an independent
Bernoulli random variable with the probability mass function
$P(c_i=1)=p_i$, $P(c_i=0)=1-p_i$, for $i=0,1,\dots,M-1$. Then the
probability distribution of $\textbf{c}$ is
\begin{equation}\label{eq7}
f(\textbf{c},\textbf{p})=\prod_{i=0}^{M-1}p_{i}^{c_i}(1-p_i)^{1-c_i}.
\end{equation}
%
%
In order to solve (\ref{eq6}) by using PMCE, we first randomize the
deterministic problem by $f(\textbf{c, p})$ for $\textbf{p}\in
[0,1]^M$ and $\textbf{c}\in\{0,1\}^M$. That is to associate
(\ref{eq6}) with  the problem of estimating the probability
$P\{F(\textbf{c})\leq\gamma\}$ for a given PAPR threshold $\gamma$.

The idea of the PMCE algorithm is to iteratively generate the
sequences $\gamma_{j}$ and $\textbf{p}_{j}$, which converge to the
optimal tuple $\gamma^{*}$ and $\textbf{p}^{*}$ in the sense of
minimal cross entropy~\cite{IEEEconf:11}. Then the optimal
$\textbf{c}^*$ can be obtained from $\textbf{p}^*$ by
$f(\textbf{c},\textbf{p})$.
More specifically, we initialize the PMCE algorithm by setting
$\textbf{p}=\textbf{p}_{0}$, and choosing a $\rho\in (0,1)$ (called
rarity parameter in PMCE~\cite{IEEEconf:11}) such that the
probability of the event $\{F(\textbf{c})\leq\gamma\}$ is around
$\rho$. Each iteration of the PMCE consists of two main phases
\cite{IEEEconf:11}:

1) For a given $\textbf{p}_{j-1}$, randomly generate a set of
samples $\textbf{c}_{1}^{j-1},\cdots,\textbf{c}_{J}^{j-1}$ from
$f(\textbf{c, p}_{j-1})$, and then calculate the PAPRs
$F(\textbf{c}_{1}^{j-1}),\cdots,F(\textbf{c}_{J}^{j-1})$. Sort
$F(\textbf{c}_{1}^{j-1}),\cdots,F(\textbf{c}_{J}^{j-1})$ in an
increasing order and denote it as
 $F_{(\it{1})}^{j-1},\cdots,F_{(\it{J})}^{j-1}$. Assign
\begin{equation}\label{yajun8}
\gamma_{j}=\frac{1}{\lceil \rho J\rceil}\sum_{k=1}^{\lceil \rho
J\rceil} F^{j-1}_{(k)},
 \end{equation}
where $\lceil \cdot\rceil$ is the ceiling function.

2) The $\textbf{p}_{j}=(p_{j,0},\cdots,p_{j,M-1})$ is updated as
\begin{equation}\label{eq9}
p_{j,i}=\frac{\sum_{k=1}^{J}I_{\{\textbf{c}_{k,i}^{j-1}=1\}}\exp{(-F(\textbf{c}_{k}^{j-1}){\lambda}_{j})}}
{\sum_{k=1}^{J}\exp{(-F(\textbf{c}_{k}^{j-1}){\lambda}_{j})}},
\end{equation}
where the indicator function $I_{\{x=1\}}=1$ if $x=1$ and $0$
otherwise, and the parameter ${\lambda}_{j}$ are obtained from the
solution of the following equation~\cite{IEEEconf:11}
\begin{equation}\label{eq10}
{\gamma}_{j}=\frac{\sum_{k=1}^{J}F(\textbf{c}_{k}^{j-1})\exp{(-F(\textbf{c}_{k}^{j-1}){\lambda}_{j})}}
{\sum_{k=1}^{J}\exp{(-F(\textbf{c}_{k}^{j-1}){\lambda}_{j})}}.
\end{equation}
In order to prevent a fast convergence to a local optimum, instead
of directly using (\ref{eq9}), we use a smoothed
version~\cite{IEEEconf:11}

\begin{equation}\label{eq11}
\hat{\textbf{p}}_{j}=\alpha\textbf{p}_{j}+(1-\alpha)\textbf{p}_{j-1},
\end{equation}
where $\alpha$ $(0<\alpha<1)$ is called a smoothing parameter.

It is important  to note that Eq.~(\ref{eq9}) is similar to the
standard CE heuristic formula $(8)$ in~\cite{IEEEconf:13}, with the
only difference that the indicator function in the CE updating
formula $I_{\{F(\textbf{c}_{k}^{j-1})\leq\gamma \}}$ is replaced by
$\exp{(-F(\textbf{c}_{k}^{j-1})\lambda}_{j})$. Eq.~(\ref{eq9}) is
preferable to the standard CE formula $(8)$ in~\cite{IEEEconf:13},
because PMCE uses the entire set of samples, whereas the standard CE
only uses the \textquotedblleft elite\textquotedblright~samples
while updating $\textbf{p}$. ~A nearly optimal solution
$\textbf{c}^{*}$ that results  in lower PAPR will be generated by
the PMCE method.

Our proposed PMCE  PAPR-reduction algorithm  can thus be summarized
as follows.\\
\begin{enumerate}\it
\item  Initialize $\hat{\textbf{p}}_0={[0.5,0.5,0.5,\dots,0.5]}$, $\rho$,
  and $\alpha$.
\item  Generate $J$ samples $\textbf{c}_1^{j-1},\dots,
  \textbf{c}_J^{j-1}$ from the density
$f(\textbf{c},\hat{\textbf{p}}_{j-1})$ and compute their PAPR
$F(\textbf{c}_k^{j-1})$ for $k=1,\cdots, J$.
\item Compute
  $\gamma_{j}$ by (\ref{yajun8}),
and use (\ref{eq10}) to find $\lambda_{j}$.

\item  Update $\textbf{p}_{j}$ by (\ref{eq9}).

\item  Obtain the smoothed $\hat{\textbf{p}}_{j}$ by (\ref{eq11}).
\item  If $\textbf{0}<\hat{\textbf{p}}_{j}<\textbf{1}$ for some $j$,
  return to step~2.
Otherwise, output the optimal solution
$\textbf{c}^*=1-2\textbf{p}^{*}$ and stop.

\end{enumerate}

\section{Simulation results}\label{sec:4}

In our simulation, quadrature PSK (QPSK)  modulation with $N=256$
sub-carriers is  used. In order to obtain  the complementary
cumulative distribution
function (CCDF) $\Pr(PAPR>PAPR_0)$, $10^5$ random OFDM symbols are
generated. The transmitted signal is oversampled by a factor of
$L=4$ for accurate PAPR~\cite{IEEEconf:10}.

\begin{figure}
\centering
\includegraphics[width=3.2in,angle=0]{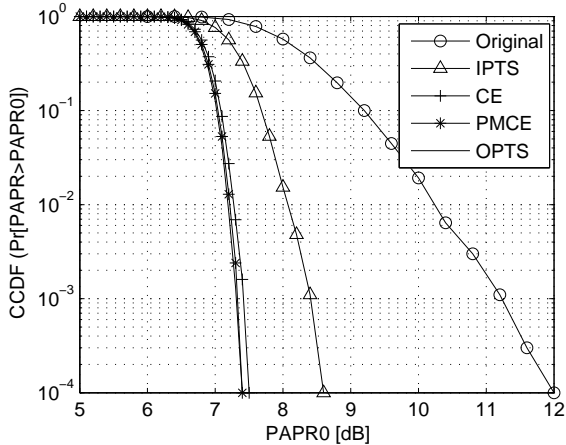}
\caption{Comparison of PAPR reduction by different methods.}
\label{fig2}
\end{figure}

In Fig.~\ref{fig2}, the CCDF for the sub-blocks of $M=8$ using
random partition is shown. In the  PMCE algorithm, $\rho=0.1$,
$\alpha=0.6$ and the sample numbers $n=40$.  When CCDF\,$=10^{-4}$,
the PAPR of the conventional  OFDM is $12$\,dB. The PAPR of  IPTS
with $(M-1) W=7\cdot 2=14$ searches  is $8.6$\,dB. The PAPRs of PMCE
and CE  with  $22$ searches are $7.4$\,dB and $7.5$\,dB
respectively. The PAPR of the optimal PTS (OPTS) with  $2^8=256$
searches  is $7.4$\,dB. Compared to the OPTS technique,  PMCE thus
offers more or less the same PAPR reduction with lower complexity
and obtains  the nearly optimal phase factors.

\begin{figure}
\centering
\includegraphics[width=3.2in,angle=0]{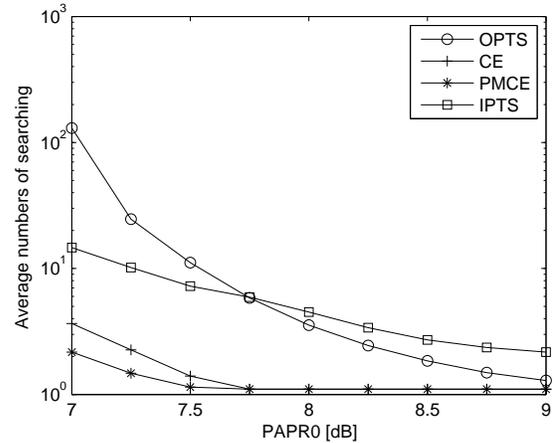}
\caption{Average numbers of searching for different methods with
thresholds.} \label{fig3}
\end{figure}

In Fig.~\ref{fig3}, we compare the average number of searchers  of
OPTS, PMCE, CE and IPTS for the thresholds $T=7,~7.25,~7.5,~7.75,~8,
~8.25,~8.5,~8.75,~9$. Here, these algorithms are terminated whenever
a phase factor that leads to a PAPR below  the threshold $T$ is
found.
Fig.~\ref{fig3} reveals  that the PMCE has lower complexity than
OPTS and IPTS for all thresholds. For the thresholds between
$7.75$\,dB and $9$\,dB, PMCE and CE has the same complexity. For the
thresholds between $7$\,dB and $7.75$\,dB, PMCE has less searching
complexity than CE. Fig.~\ref{fig3} shows that PMCE achieves a low
PAPR and decreases the computational complexity.

\section{Conclusion}
In this paper, we propose a  PMCE-based PTS algorithm. The
algorithm finds a nearly optimal combination of phase factors for
OFDM signals, with significantly reduced computational complexity.
Simulation results show that our method outperforms the existing
methods both in the CCDF of PAPR and the computational complexity.



%


%
%

\end{document}